\documentclass[%
 aip,
 amsmath,amssymb,
 reprint,%
]{revtex4-1}
\usepackage{graphicx}
\usepackage{dcolumn}
\usepackage{bm}

\usepackage[utf8]{inputenc}
\usepackage[T1]{fontenc}
\usepackage{mathptmx}
\usepackage{mhchem}
\draft 
\usepackage{soul,color}
\soulregister\cite7
\soulregister\ref7
\soulregister\pageref7

\begin{document}

\preprint{AIP/123-QED}
\title{Fabrication of Ultra-High Q Silica Microdisk Using Chemo-Mechanical Polishing} 



\author{S. Honari}
\author{S. Haque}

\author{T.Lu}
 \email{taolu@ece.uvic.ca}
 
\affiliation{%
Electrical and Computer Engineering Department, University of Victoria, \\EOW 448, 3800 Finnerty Rd., Victoria, BC V8P 5C2, Canada
}%

\date{\today}

\begin{abstract}
Here we demonstrate that adding a chemo-mechanical polishing (CMP) procedure to conventional photolithography, a silica microdisk with ultra-high quality factors ($>10^8$) can be fabricated. By comparing with the intrinsic optical quality factor (Q) measured at 970~nm, we observe that due to the significantly reduced surface roughness, at 1550~nm wavelength the water molecule absorption at the cavity surface supersedes Rayleigh scattering as the dominant factor for Q degradation. 
\end{abstract}


\maketitle 

The study of whispering gallery mode~(WGM) micro resonators has dominated many fields of science over past decades. Microcavities have become one of the most attractive optical components to study in a broad range of scientific disciplines, including but not limited to, nanoparticle detection, biosensing, quantum information, comb generation, and optomechanics~\cite{vahala2003optical,lu2011high,baaske2014single,kippenberg2011microresonator,papp2014microresonator,kippenberg2008cavity}. The power of microcavities to confine the light is mostly measured by optical quality factor (Q) as the higher the Q goes the more intense the light inside the cavity becomes.~\cite{campillo1991cavity} High Q microcavities are essential for better detection in sensing applications, lower threshold power for comb generation, and parametric oscillation~\cite{kippenberg2004kerr,ilchenko2004nonlinear} and narrower linewidth for laser oscillators.~\cite{li2012characterization}

 One of the most important limiting factors for achieving high Q is Rayleigh scattering induced by surface roughness~\cite{gorodetsky2000rayleigh}. Hence, efforts to increase the optical Q has mainly relied on using the expensive fabrication techniques such as stepper lithography~\cite{lee2012chemically} or \ce{CO_2} laser reflow~\cite{armani2003ultra} to decrease the surface roughness of microcavities and increase the Q. However, these techniques come with either the need for expensive machinery or incompatibility with monolithic integration. Silica microdisks with Q exceeding 1 billion have been fabricated recently by carving out the Silica and making trenches on grown silica film rather that making stand alone disks.~\cite{wu2020greater} High Q microresonators and low loss waveguides from other nonlinear materials have also been demonstrated with deposition on silica platforms to take advantage of smooth surface and low loss nature of the oxide.~\cite{kim2020universal} 
 
Chemo-mechanical polishing (CMP) has been used  to increase the optical Q for \ce{LiNbO_3}~\cite{Wu_Lithium,wang2019chemo} and \ce{Si_3N_4}~\cite{ji2017ultra} microdisks, but to the best of our knowledge, this technique has not been used to fabricate \ce{SiO_2} microdisks.
In this paper we incorporate CMP to conventional photolithography to fabricate ultra-high Q microdisks. Through scanning electron microscopy (SEM) and numerical modelling, we further analyze limiting factors to further Q improvement.


\begin{figure}
 \includegraphics[width=0.5\textwidth]{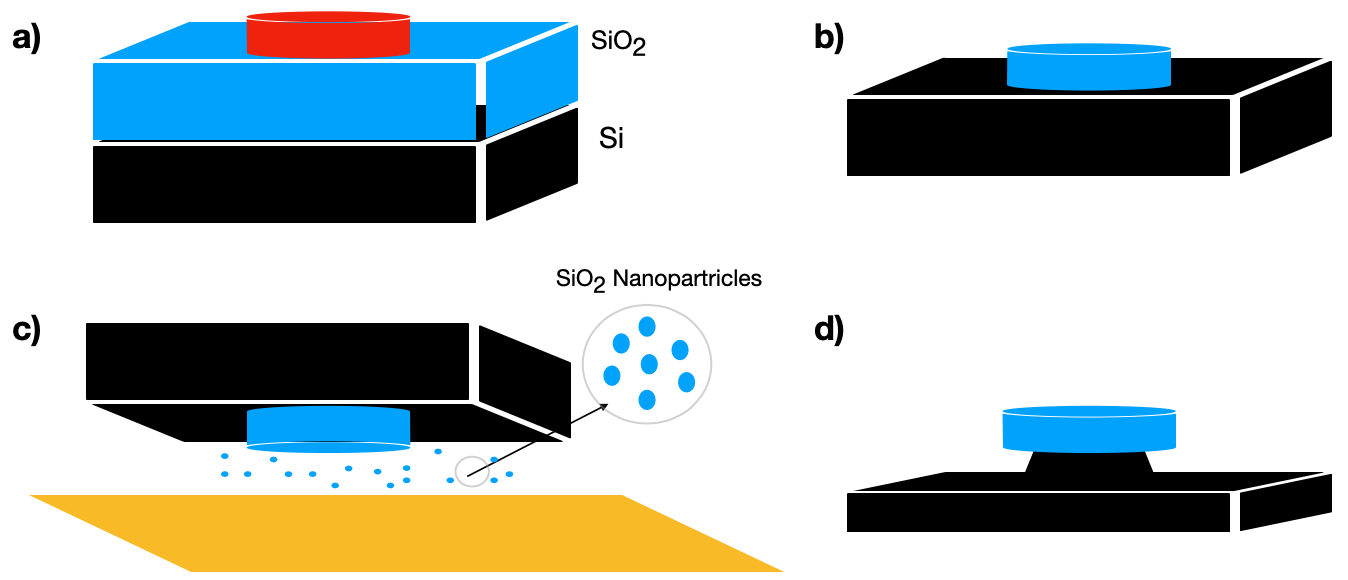}
\caption{a) Photolithograhpy of the resist to transfer the pattern, b) HF wet etch to make the disk, c) CMP step to increase the smoothness of the surface, and d) etching Si substrate to make the pillar.}
\label{fig:fabricationsteps}
 \end{figure}
Our fabrication process consists of four steps: a) conventional photolithography using deep UV light, b) wet etching of \ce{SiO_2} using buffered \ce{HF}, c) chemomechanical polishing of the \ce{SiO_2} disks with silica slurry, and d) \ce{XeF_2} dry etch of the Silicon layer to form the pillars (Fig~\ref{fig:fabricationsteps}).

The fabrication details are as follows. After applying a standard wafer cleaning process, we spin coat a silica-on-silicon wafer (SoS, University Wafers) with positive resist (S1813). A UV lithography is then applied to transfer the microdisk pattern from the photomask onto the wafer followed by a buffered HF (Transene) wet etch to form disks on the silica thin film.  To further improve the surface smoothness of the disks, we proceed to the polishing step using slurry of 70-nm-diameter silica nanoparticles. Before the polishing step can be done, it is important to change the surface chemistry of the sample. Hydrogen passivation surface of the sample after HF wet etching makes the surface extremely hydrophobic,~\cite{trucks1990mechanism} which can hinder the flow of slurry and reduce the effectiveness of the CMP step significantly. In order to remedy this problem, we used a 10 minutes Piranha solution~(\ce{H_2SO_4}/\ce{H_2O_2}) to clean the surface and make it hydrophylic. Hand polishing of the sample was then performed using the silica nanoparticle slurry to reduce the surface roughness. RCA cleaning steps has also been performed after polishing to wash off the slurry particles stick to the surface of the disk. In the last step of fabrication, \ce{XeF_2} gas was used inside an etching chamber to undercut the disk and form the pillars by etching the silicon underneath the disks. This step is crucial for achieving high Q micro resonators, since the light silicon interaction inside the cavity should be minimized. The maximum amount of undercut is limited by the buckling effect that happens at larger undercuts, and can degrade the Q or even make the disk crack. However, the problem can be mitigated by high temperature annealing (1,000~$^\circ{\rm C}$).~\cite{wu2020greater}
\begin{figure*}[ht]
    \centering
    \includegraphics[width=0.9\textwidth]{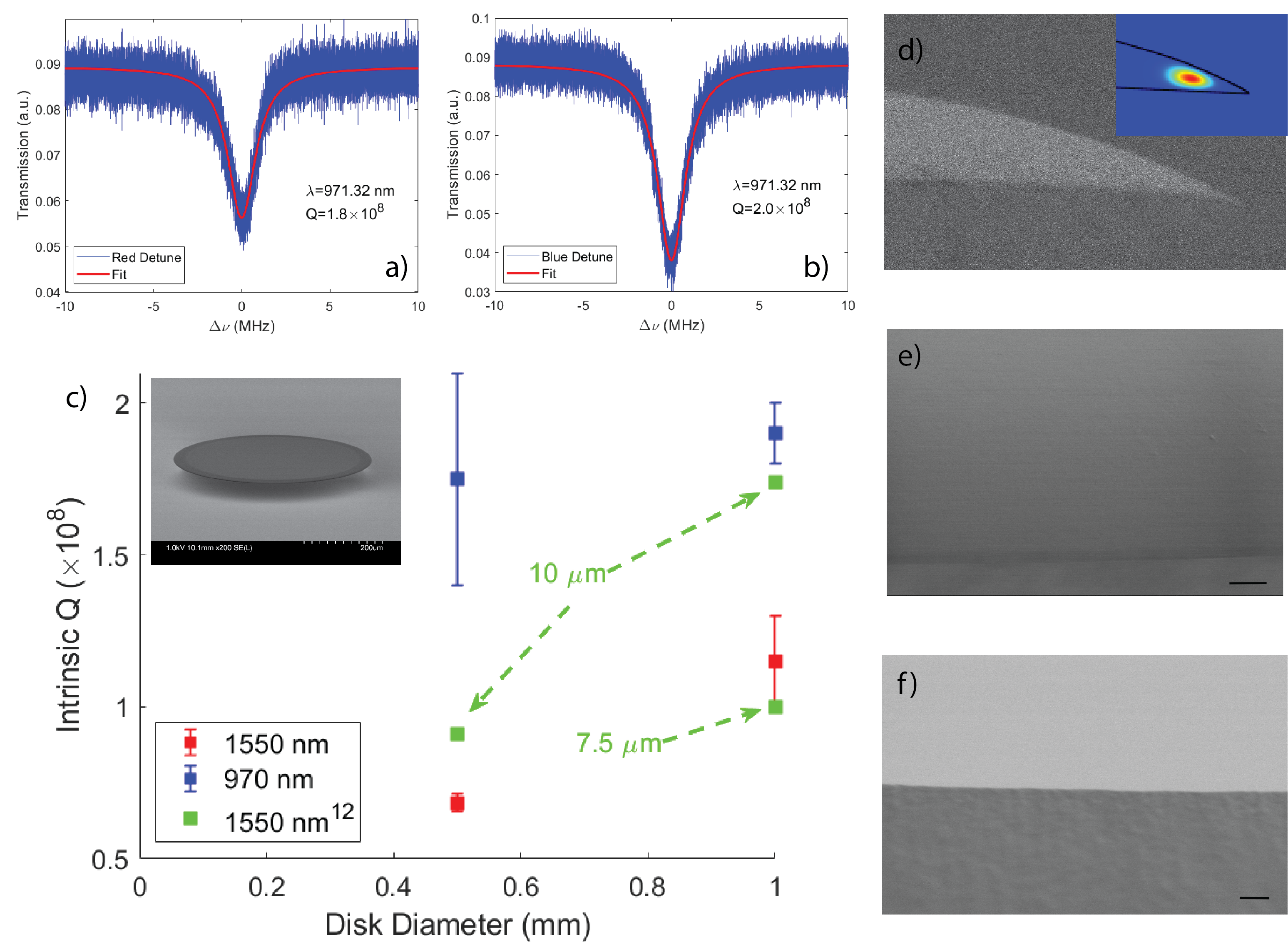}
    \caption{Transmission spectra of a 1~mm disk when the 970~nm probe laser wavelengths scanned toward a) red (blue traces) and b) blue detune (blue trace). The intrinsic Qs of $1.8\times10^8$ and $2.0\times10^8$ were obtained through least square fitting to a Lorentzian function (red traces) respectively, indicating the intrinsic Q to be $(1.9\pm0.1)\times10^8$. c) Intrinsic Q as a function of disk diameter at 970~nm (blue squares), 1550~nm (red squares) wavelength bands with the lengths of the error bars corresponding to the differences of the measured intrinsic Q at red and blue detune. In comparison, the recorded values of 7.5~${\rm \mu}$m and 10~${\rm \mu}$m thick microdisks by Lee et. al.~\cite{lee2012chemically} are displayed as green squares. Note that, after polishing, the silica thin film thickness of our disks reduced from 10~${\rm \mu}$m to 7.5~${\rm \mu}$m. d) The side view SEM micrograph of a polished disk at its edge shows the side wall vanishes after polishing. e) the topview micrograph further confirms the sharp corner at the edge of the upper surface were polished away as oppose to the unpolished disk displayed in f). The comparison also demonstrates the reduced surface roughness through CMP.}
    \label{fig_sem_and_Q_data}
\end{figure*}

In the next step, we  measure the intrinsic Qs at 970~nm and 1550~nm wavelength bands. Here, external cavity tunable lasers (Samtec TSL-550 for 1550~nm, and Newfocus TLB 6718-D for 970~nm) were used to probe the microdisks through a tapered optical fiber. As shown in Fig.~\ref{fig_sem_and_Q_data}a, by linearly scanning the laser around the cavity resonance wavelength, a Lorentzian shaped signal will appear at the output transmission of the tapered fiber, from which intrinsic Q can be calculated. Here, to accurately readout the optical frequency, we placed a fiber Mach-Zehnder interferometer in parallel to the microdisk cavity as a reference.~\cite{lu2011high} Note that,the intrinsic Q values measured by red detuning and blue detuning of the laser wavelengths are slightly different, due to the thermal effects. Therefore, we take the average of both Qs as the estimated value of intrinsic Q and their difference as the uncertainty. For example, as shown on Fig.~\ref{fig_sem_and_Q_data}a, the Q measured from blue detuning on a 10~${\rm \mu}$m (7.5~${\rm \mu}$m after polishing) thick, 1mm diameter microdisk using a 970~nm laser was $2.0{\times}10^8$ while Fig.~\ref{fig_sem_and_Q_data}b shows that from a red detuned measurement the Q of $1.8{\times}10^8$ was obtained. Consequently we determine the intrinsic Q to be $(1.9\pm0.1){\times}10^8$ as shown in Fig.~\ref{fig_sem_and_Q_data}c (blue square with error bars representing the uncertainty). We also used our fabrication process to make 500~${\rm \mu}$m disks, and the average intrinsic Q of $(1.8\pm0.4)\times{10^8}$ was obtained. In order to better compare our work with previously reported results (green squares),~\cite{lee2012chemically} we used 1550~nm laser to measure the Q as well. We observed the optical Q drops to $(1.2\pm0.2){\times}10^8$ for 1~mm disk and $(6.9\pm0.2){\times}10^7$ for 500~${\rm \mu}$m disk~(red squares in Fig.~\ref{fig_sem_and_Q_data}b). Note that the Q of our 1~mm disk is slightly better than the equivalent 7.5~${\rm \mu}$m thick disks reported in.~\cite{lee2012chemically}
The side view and top view SEM micrographs in Fig.~\ref{fig_sem_and_Q_data}d-e show reduced roughness on the polished surface compared to the unpolished one shown in Fig.~\ref{fig_sem_and_Q_data}f and confirms our process make the disk edges more smooth, which will reduce the Rayleigh scattering. The edge shape has altered after the CMP step and the sharp angle between the top surface and the sidewall has changed to a smooth gradual change in angle, so that top surface and the sidewall are almost indistinguishable. This change in turn would affect the mode position inside the cavity as shown in the inset of Fig.~\ref{fig_sem_and_Q_data}d.

Further, the observation of lower Q at longer wavelength suggests Rayleigh scattering is not the dominant factor for Q degradation. Instead, water absorption at the disk surface dominates the Q at 1550~nm since otherwise surface roughness induced Rayleigh scattering should make the Q at shorter wavelength (970~nm) even lower.~\cite{borselli2004rayleigh} This contradicts the previous observations that Q at 1550~nm is limited by surface roughness.~\cite{vernooy1998high,rokhsari2004loss,ganta2014measuring} To confirm, we numerically simulate the modes and compare the intrinsic Q at both wavelengths. In simulation, the side view SEM image of the disk in Fig.~\ref{fig_sem_and_Q_data}d was converted to a contour curve for COMSOL simulation as shown on its inset (black trace).  In the absence of water layer, the Q of $2.12{\times}10^{10}$ at 970~nm and $2.69\times10^{10}$ at 1550~nm were obtained through simulations. Since our SEM does not provide sufficient resolution for surface roughness, the simulated Q were mainly silica material absorption limited. We then added a layer of water at the disk surface and obtained a Q of $1.56{\times}10^{8}$ at 1550~nm when the water layer is 1~nm thick. This is in close agreement to our experiment observation around this wavelength. On the other hand, the simulated Q at 970~nm was $4.14\times10^9$ due to the significantly lower water absorption at this wavelength band. As the experimental measurement (Q=(1.9$\pm$0.1$)\times10^{8}$) is much lower than the simulation value, we confirm that at 970~nm Q is still limited by Rayleigh scattering. In fact, assuming the Rayleigh scattering induced Q (${\rm Q}_{ss}$) has $\lambda^3$ dependence~\cite{borselli2004rayleigh} and ${\rm Q}_{ss}\sim1.9\times10^8$ at 970~nm, one would expect ${\rm Q}_{ss}\sim{7.8\times10^{8}}$ at 1550~nm, which is much higher than the measured Q. From these observations, we concluded that at 1550~nm, Q is limited by a 1-nm water layer absorption while at 970~nm surface roughness is still the limiting factor for Q.

In summary, we made microdisks with ultra high optical Q using conventional photolithography in conjunction with CMP. As a result, surface roughness is no longer a limiting factor for Q at 1550~nm. The Q at shorter wavelength can be further improved by using smaller nanoparticles as slurry to achieve lower surface roughness. Through simulation, we further find that by removing the 1~nm water layer at the microdisk surface, a Q value close to $10^9$ can be reached at 1550~nm.  

 This work was supported in part by the Nature Science and Engineering Research Council of Canada (NSERC) Discovery (Grant No. RGPIN-2020-05938),  and Threat Reduction Agency (DTRA) Thrust Area 7, Topic G18 (Grant No.GRANT12500317). We would like to acknowledge CMC Microsystems for the provision of products and services that facilitated this research, including the use of COMSOL for the numerical analysis. The authors would like to thank Dr. Elaine Humphrey, Mr. Jon Rudge, and the staff of Advanced Microscopy Facility (AMF) at University of Victoria for their help and constructive and valuable discussions. The data that support the findings of this study are available from the corresponding author upon reasonable request.
%

\end{document}